\documentclass[
 reprint,
 superscriptaddress,
 amsmath, amssymb,
 showkeys,
 aps,
 pre,
]{revtex4-2}

\usepackage{graphicx}          
\usepackage{hyperref}          
\usepackage{xcolor}           

\graphicspath{{figs/}}		

\begin{document}

\title{Self-Repelling Bi-Exploration Process}

\author{H. Dashti N.}
\email{hdashti@kias.re.kr}
\affiliation{School of Physics, Korea Institute for Advanced Study, Seoul 02455, Korea}

\author{M. N. Najafi}
\email{morteza.nattagh@gmail.com}
\affiliation{Department of Physics, University of Mohaghegh Ardabili, P.O. Box 179, Ardabil, Iran}

\author{Hyunggyu Park}
\email{hgpark@kias.re.kr}
\affiliation{School of Physics, Korea Institute for Advanced Study, Seoul 02455, Korea}


\begin{abstract}
Self-repelling two-leg (biped) spider walk is considered where the local stochastic movements are governed by two independent control parameters $ \beta_d$ and $ \beta_h $, so that the former controls the distance ($ d $) between the legs positions, and the latter controls the statistics of self-crossing of the traversed paths. The probability measure for local movements is supposed to be the one for the ``true self-avoiding walk'' multiplied by a factor exponentially decaying with $ d $. After a transient behavior for short times, a variety of behaviors have been observed for large times depending on the value of $\beta_d$ and $\beta_h$. Our statistical analysis reveals that the system undergoes a crossover between two (small and large $\beta_d$) regimes identified in large times ($t$). In the small $\beta_d$ regime, the random walkers (identified by the position of the legs of the spider) remain on average in a fixed non-zero distance in the large time limit, whereas in the second regime (large $\beta_d$s), the absorbing force between the walkers dominates the other stochastic forces. In the latter regime, $ d $ decays in a power-law fashion with the logarithm of time. When the system is mapped to a growth process (represented by a height field which is identified by the number of visits for each point), the roughness and the average height show different behaviors in two regimes, i.e., they show power-law with respect to $t$ in the first regime, and $\log t$ in the second regime. The fractal dimension of the random walker traces and the winding angle are shown to consistently undergo a similar crossover.
\end{abstract}


\maketitle


\section{Introduction}

Random walks are in the heart of non-equilibrium statistical mechanics and stochastic processes. They are unique in describing nature due to their large applications in plenty of physical systems, like the polymers in a good solvent \cite{de1979scaling}, the stock markets \cite{osborne1959brownian}, the thermal motion of gas molecules \cite{tsekov2010brownian} or networks \cite{avin2006power, boyd2005mixing}, and  mathematical statistics \cite{spitzer2013principles}.
In most cases, the studies on random walks in the literature are surprisingly limited to a few cases like uncorrelated random walks, self-avoiding and loop-erased random walks \cite{lawler1980self, rudnick1987winding, lawler1999loop, bauer2008lerw,sepehrinia2019random}, and fractional Brownian motions \cite{beran1994statistics} for which the mathematical structures are more or less known. Many properties of these random walks in various dimensions have been calculated analytically and numerically. In two dimensions, we know that self-avoiding walk (SAW) is a Schramm-Loewner evolution (SLE) with a diffusivity parameter $\kappa=\frac{8}{3}$ which is consistent with a conformal field theory (CFT) with a central charge $c=0$ \cite{cardy2005sle, saberi2010class}, whereas loop-erased random walk (LERW) is described by SLE$_{\kappa=2}$ \cite{schramm2000scaling}, which is consistent with the $c=-2$ CFT \cite{majumdar1992equivalenc} (both CFTs are logarithmic). The former reveals a relation with the critical percolation theory \cite{mathieu2007percolation}, and the latter shows that LERW is consistent with the interfaces of sandpiles \cite{Saberi2009Direct}. Some authors occasionally consider more sophisticated situations, like correlated random walks with dropping debris (namely true SAW, TSAW) \cite{amit1983asymptotic}, self-avoiding random walks in a media with quenched randomness \cite{grassberger1993recursive}, TSAW with diffusion of debris \cite{grassberger2017self}, LERW in the correlated background \cite{cheragh2019corr}, and random walks on the random graphs \cite{boyd2005mixing, craswell2007random}. Another type of correlation is the one that the motion of a random walker depends on the effective environment that is formed by the rest of the random walkers in the media. This problem applies to many systems like the active matter (like the Vicsek model of self-propelled particles \cite{chate2008modeling} and the active Brownian motion \cite{volpe2014simulation}), thermal motion of gas molecules \cite{tsekov2010brownian}. Polymers \cite{fisher1966shape}, polymer brushes \cite{milner1991polymer}, and the trace of grains in sandpiles \cite{najafi2020invasion} are other examples.

In nature, there are some more sophisticated situations, like multi-agent stochastic correlated walks, which can serve as the example of few body active dynamics, taking conditional steps depending on the structure of the background potential or effective interaction with other agents. Consider as an example two (male and female) insects that besides seeking food, intend to each other, and therefore perform correlated exploration process in two dimensions, with a low tendency to step on the places that they have already stepped on due to the fact that the chance of finding food in the traversed path is low. This problem can be considered as a combination of TSAW and multi-agent random walk problem, which we call \emph{self-repelling bi-exploration process} (SRBP). SRBP can be taken into account as a member of a larger class, namely the \textit{Spider walks}, defining the systems in which the particles move in such a way that their movements do not violate some given rules~\cite{gallesco2011note}. DNA molecular biped on a one-dimensional walking path is another example that is mapped to spider walk on a one-dimensional~\cite{antal2007molecular,antal2007molecular0,ben2011stochastic} and two-dimensional ~\cite{antal2012molecular} random media. There are many more examples that can be mapped to our model (SRBP as a generalization of biped spider walk) like insect movement~\cite{kareiva1983analyzing}, polymer ring entangled with obstacles~\cite{grosberg2003winding}, local clustering for multi-agent random walks~\cite{alamgir2010multi}, and animal's movement as correlated random walks~\cite{bovet1988spatial}. Another example of the systems that can potentially be mapped to SRBP is a system with two kinds of monomers (say blue and red) with an absorbing interaction between blue-red pairs and repulsion between blue-blue and red-red pairs, combining of which two (blue and red) self-avoiding polymers are constructed, which serves as a generalization of polymers entangled with obstacles~\cite{grosberg2003winding}.

In this paper, we consider the SRBP problem with two independent parameters, one of which controls the tendency between two agents ($ \beta_d \equiv 1 / T_d $), and another controls the disinclination for crossing the traversed path ($ \beta_h \equiv 1 / T_h $). The variables $T_d$ and $T_h$ can be interpreted as two different temperatures in our model. To capture the ``true self-repulsion", we use the method given in \cite{grassberger2017self}, according to which the random walkers drop one unit of debris in the site that they are in, so that $ h_i(t) $ shows the height of the debris in site $i$ at time $t$. Then, the random walkers come back to any site $i$ with a probability proportional to $\exp\left[ -\beta_h h \right] $ (the step length is one unit of lattice). The relative distance of two agents, the height of debris, and the random walker paths are the important quantities that we study in this paper. This system is shown to undergo anomalous diffusion (with respect to the relative coordinate) and show a crossover point to a new phase that is determined by $\beta_d$ and $\beta_h$.

The paper is organized as follows: In the next section, we introduce the model. The results for the diffusion process are presented in Sec.~\ref{SEC:diff}. Sec.~\ref{SEC:FD} is devoted to the fractal dimension of the traces and the winding angle statistics. We close the paper with a conclusion section.

\section{The Model}

The spider walks with $k$ legs are defined through considering $k$ different (coupled) traces ($X_t=(X_{1,t},X_{2,t},...,X_{k,t})$ where $X_{i,t}$ stands for the position of the $i$th leg of the spider at time $t$) over a given undirected connected graph $G(V,E)$ with vertex set $V$ and edge set $E$. The model is identified using the transition matrix $P=\left\lbrace p(x,y)\right\rbrace_{x,y\in G} $, where $p(x,y)$ is zero only when the required links are missing in $G$. Showing the position of the spider by $\textbf{x}=(x_1,x_2,...,x_k)$, the transition to $\textbf{y}=(y_1,y_2,...,y_k)$ is given by $p(x_i,y_i)$ \textit{if there exists exactly one index $i$ such that $x_i\ne y_i$}. Many properties of this model have been explored in the literature, like recurrence~\cite{gallesco2011note}, transience, ergodicity, spider walk in the random media~\cite{takhistova2017spider}. The example is the legs of the biped molecule (as a biped spider) which moves on the integer lattice representing the nucleic acid binding domains imprinted on the path~\cite{antal2007molecular}.

As partially explained in the introduction, we consider two correlated random walks that step on a lattice. This problem is mapped to a generalized biped spider walk problem in the Euclidean space (square lattice). The generalization backs to the fact that the traces that traversed by the legs of the spider matter, i.e. the traces are self-repulsive in the sense that in each time step $t$, the random walkers drop a unit of debris at the point that they stand on, say the site $i$, so that the height of the site increases by one, i.e. $h_i(t)\rightarrow h_i(t)+1$. The steps are taken according to the following update probability: Suppose that the random walkers are in points $\textbf{r}^0_1$ and $\textbf{r}^0_2$ at time $t$, and $\textbf{r}_1$ and $\textbf{r}_2$ are some random neighbors of $\textbf{r}^0_1$ and $\textbf{r}^0_2$, respectively. Then the probability to step to the neighboring sites $\textbf{r}_1$ and $\textbf{r}_2$ at the next time is proportional to:
\begin{equation}
    P\propto \exp\left[ -\beta_h \left(h(\textbf{r}_1)+h(\textbf{r}_2)\right) \right] \exp \left[-\beta_d \delta d \right],
\label{Eq:P}
\end{equation}
where $\delta d\equiv d^{\text{new}} - d^{\text{old}}$, $d^{\text{new}} \equiv \left|\textbf{r}_1 - \textbf{r}_2\right|$ and $d^{\text{old}} \equiv \left|\textbf{r}^0_1 - \textbf{r}^0_2\right|$. The first factor cares about self repulsion and the second one cares about the tendency between the pair, so that when both $\beta_h$ and $\beta_d$ are zero, all the directions are equiprobable and one retrieves two-dimensional uncorrelated random walks. The simulation is started by two agents that start from the origin. At each time the next step is taken towards a random neighbor according to the probability given above. For calculating the winding angle statistics, we prevent the agents to enter a region in a close neighborhood of the origin~\cite{belisle1991winding}. The larger amount of $\beta_h$ leads to a smaller probability of self intersection, so that $(\beta_d,\beta_h)\rightarrow(0,\infty)$ gives two independent ordinary SAW. In the opposite limit for $\beta_h=0$, and defining $d\equiv \left|\textbf{r}_1-\textbf{r}_2\right|$, at long enough times, one expects that
\begin{equation}
    \left\langle d\right\rangle=-\partial/\partial \beta_d\ln \int_0^{\infty}e^{-\beta_d d}\text{d}d=\beta_d^{-1}=T_d,
\label{Eq:mean}
\end{equation}
where the ergodicity was considered, meaning that the random walker has enough time to find any possible configuration, i.e., all $d$ values are \textit{visited}. The other famous limit is $\beta_d=\beta_h=0$, which is corresponds to two independent 2D uncorrelated random walk --- space filling with mass fractal dimension $d_f=2$.\\

\begin{figure}
    \centerline{\includegraphics[scale=0.75]{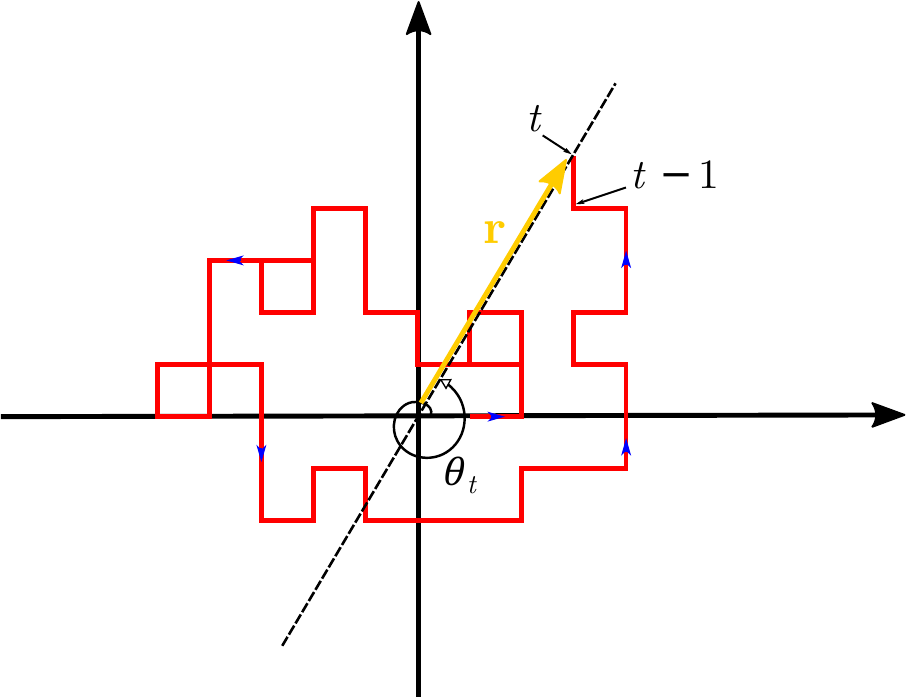}}
    \caption{Schematic representation of the random walks of one agent with the corresponding winding angle $ \theta_t $ and displacement $ \textbf{r}(t) $.}
\label{fig:winding_angle_def}
\end{figure}
The winding angle $\theta$ is defined as the total winding angle of the random walker around the origin. The general setup of the problem and the quantities of interest are schematically shown in Fig.~\ref{fig:winding_angle_def}, where a small region around the starting point was removed. It was shown that for the ordinary 2D random walks with the starting point excluded \cite{rudnick1987winding, belisle1991winding}
\begin{equation}
    \left\langle \theta^{m}\right\rangle \propto (\log t)^{2m},
\label{Eq:winding}
\end{equation}
and the distribution function $P_t(\theta)\propto \exp\left[ -2\pi |\theta|\log t \right] $. This is in contrast to self-avoiding walks (SAW), where $ \left\langle \theta^2 \right\rangle = \frac{8}{3} t $ \cite{cardy2005sle}. Also, note that the characteristic distance of the random walker scales with $t^{\nu}$, where $\nu=1/2$ for ordinary random walks and $\nu=3/4$ for SAW.

\subsection*{Summary of Results}

To improve the flow of the paper, we declare here the important findings of the paper. The significant finding of the present paper is the crossover behavior in terms of $\beta_d$. This crossover region is $\beta_d^*\in [0.2-0.5]$. \\
Figs.~\ref{fig:r_and_alpha}c and d show the exponent $\nu$ in terms of $\beta_d$ and $\beta_h$ which is defined by $r\propto t^{\nu}$. This exponent changes abruptly in the crossover region, i.e. from $\nu\approx 0.5$ for $ \beta_d \lesssim  0.2 $ to another value for $ \beta_d \gtrsim  0.2 $. \\
The difference between two regimes becomes more clear when one focuses on the time-dependence of $d$ (Figure.~\ref{fig:d_and_collapse}a and b). $d(t)$ decays with a power-law fashion with $(\log t)^{-\alpha_d}$ for $\beta_d > 0.5$, whereas for $\beta_d < 0.5$ the random walkers remain in a non-zero average distance.

In the second part of the paper, we map the system to a $(2+1)$-dimensional growth process, represented by the height configurations. The statistics of the height and other related quantities change from one regime to the other. As an example, the width $w(t)$ scales with time as $t^{\alpha_w^{(1)}}$ for $\beta_d \ll 0.2$ (Fig.~\ref{fig:w_and_alpha}a), while it follows the relation $w(t)\sim (\log t)^{\alpha_w^{(2)}}$ for $\beta_d \gg 0.2$ as depicted in Fig.~\ref{fig:w_and_alpha}b.

This change is also seen in the fractal dimension of the trace of random walkers (Fig.~\ref{fig:sandbox_df}d), i.e. $d_f$ changes abruptly from $ \beta_d \lesssim  0.2 $. to other values in the $\beta_d \gtrsim 0.2$ regime. The same change of behavior is seen for the winding angle as well (Fig.~\ref{fig:theta_and_alpha}d), where the exponent $\alpha_\theta$ changes abruptly between the regimes.

\section{The diffusion process}
\label{SEC:diff}

In this section, we present the results of the simulations. We use the Metropolis algorithm with the accept ratio $P$ defined in Eq.~\eqref{Eq:P} to accept one of the 16 possible pair movements on a lattice at each time step $ t $. For various values of $ \beta_d$, and $\beta_h$ (with variable increments), we generated more than $10^5$ independent realizations, for each of which the time runs up to $t=10^6$. The run time for high $\beta_h$ values increases dramatically because of the self-avoiding character of the traces.\\

\begin{figure*}
    \centering
    \includegraphics[width=0.9\textwidth]{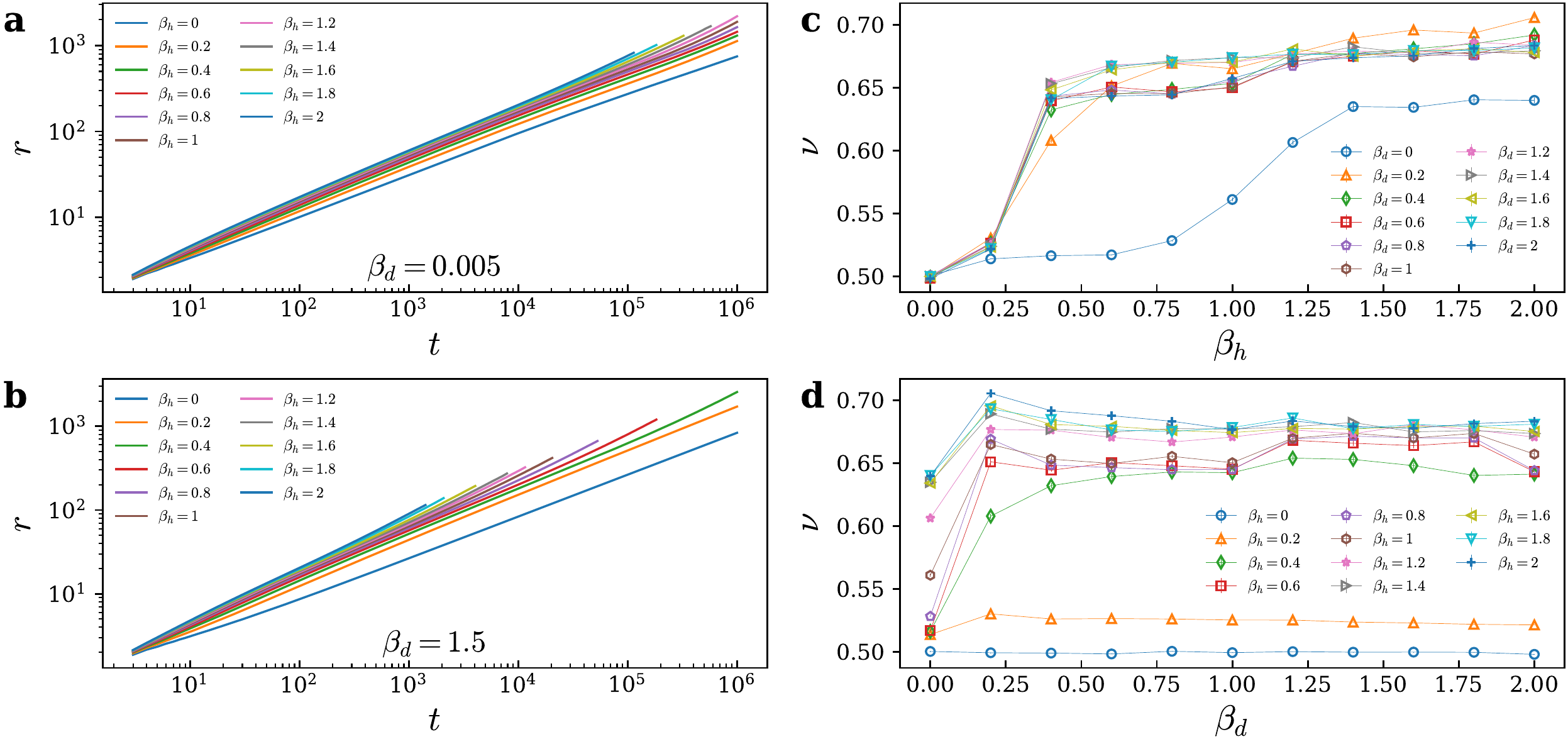}
    \caption{ Log-log plot of $r\equiv |\textbf{r}|$ versus $t$ for  $\beta_d=0.005$ and $\beta_d=1.5$ are shown in panel-a and -b, respectively. The slope of curves gives the diffusion exponent $\nu$. The corresponding exponents in terms of $\beta_h$ and $\beta_d$ are shown in panel-c and -d, respectively. In panel-d, one can see that for $\beta_d \lesssim 0.2$, the diffusion exponent $\nu$ changes abruptly, and for $\beta_d \gtrsim 0.2$, roughly remains fixed.
    }
\label{fig:r_and_alpha}
\end{figure*}

The type of diffusion (normal-, sub-, and super-diffusion) for each agent is arguably the most important question in the transport perspective. Our inspections show that the statistics of the random walkers are quite sensitive to $\beta_d$ and $\beta_h$. Importantly, the position of each random walker $r\equiv |\textbf{r}|$ crosses over from normal diffusion (identified by an exponent $\nu=\frac{1}{2}$ in the scaling relation $r\propto t^{\nu}$) to a regime with different diffusion exponent, see Fig.~\ref{fig:r_and_alpha}. As $\beta_h$ increases, one expects that the SAW behavior is retrieved, i.e., $\nu^{\text{SAW}}=\frac{3}{4}$ \cite{cardy2005sle}, which is expected from Fig.~\ref{fig:r_and_alpha}c, while the dependence on $\beta_d$ is quite low for $\beta_d\gtrsim 0.2$. As $\beta_d$ increases, the crossover to the new regime happens earlier, i.e., $\beta_d$ facilitates this crossover.

\begin{figure*}
    \centering
    \includegraphics[width=0.9\textwidth]{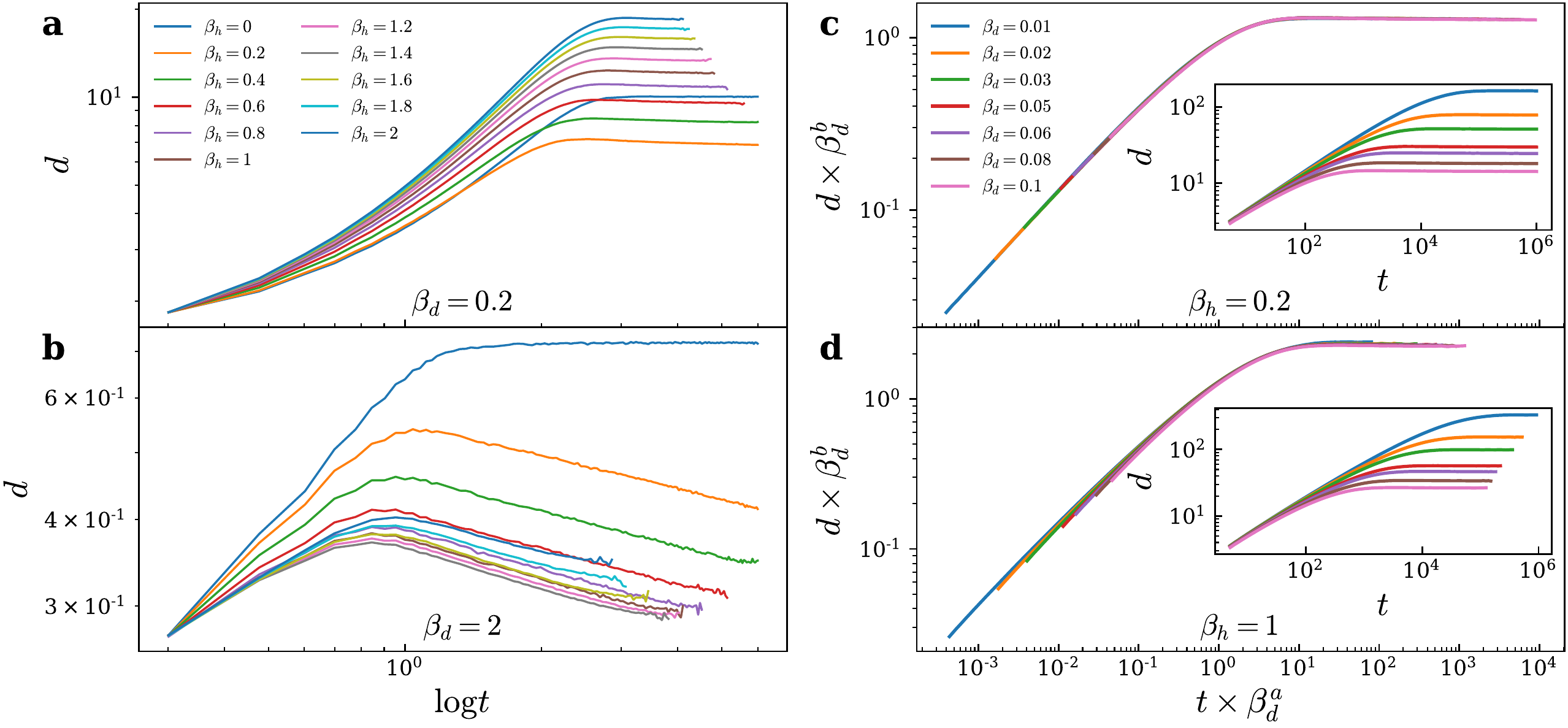}
    \caption{ Log-log plot of $d\equiv |\textbf{r}_1-\textbf{r}_2|$ in terms of $\log t$ for $\beta_d=0.2$ and for $\beta_d=2$ are plotted in panel-a and -b, respectively.
    The legend in panel-b is the same as in panel-a. It is clear that we have two different behaviors in panel-a and -b. The $\beta_d = 0.2$ and $\beta_d=2$ belong to R1 and R2 regimes, respectively. When $\beta_h \approx 0$, these differences do not exist anymore; i.e., the variable $d(t)$ reaches a saturation value $d_s$ at a big enough time for all values of $\beta_d > 0$. The corresponding data collapse analysis for $\beta_h=0.2$ and $ \beta_h=1 $ with $a=2.00(5)$ and $b=1.00(5)$ are shown in panel-c, and -d, respectively. It is worth noting that the data collapse happens in the R1 regime ($\beta_h > 0$, $\beta_d \lesssim 0.2$).
    }
\label{fig:d_and_collapse}
\end{figure*}

The relative distance between the random walkers ($d_{\beta_d,\beta_h}(t)$) is the other quantity that shows considerable change as $\beta_d$ and $\beta_h$ vary. Figure~\ref{fig:d_and_collapse} shows this quantity in terms of time $t$ for various amounts of $\beta_h$ and for $\beta_d=0.2$ and $2$ (a and b respectively). For both cases in the early times, the distance between the random walkers (agents) increases with time in a power-law fashion. There is however an important difference between them in long times, i.e., for $\beta_d=0.2$ the graph saturates to a $\beta_h$-dependent constant, while the graph for $\beta_d=2$, $d$ decays in a power-law fashion\textit{ in terms of $\log t$}. More precisely, in the large $\beta_d$ regime, the random walkers are asymptotically absorbed to each other with a heavy tail function
\begin{equation}
	\left. d_{\beta_d,\beta_h}^{\text{R2}}(t)\right|_{\text{large times}}\propto \left(\log t \right) ^{-\alpha_d}.
\label{Eq:R2-Scaling}
\end{equation}
Our observations show that a crossover is established between two distinct regimes in terms of $\beta_d$ identified by different statistical behaviors. Let us show the crossover region by $\beta_d^*$ which is $[0.2, 0.5]$. For $\beta_d$s smaller than $\beta_d^*$ (let's call it R1 regime), $d$ saturates to a constant value for long enough times (like $\beta_d=0.2$ in Fig.~\ref{fig:d_and_collapse}a), while for $\beta_d>\beta_d^*$ (R2 regime) $d$ varies in the form of Eq.~\ref{Eq:R2-Scaling} (like $\beta_d=2$ in Fig.~\ref{fig:d_and_collapse}b). For R1 regime, the curves for $d$ are collapsed (fitted to each other) with an appropriate choice of exponents, the fact that was not observed for R2 regime. Figures~\ref{fig:d_and_collapse}c and~\ref{fig:d_and_collapse}d show the data collapse analysis for the R1 regime for $\beta_h=0.2$ and $1$, demonstrating that the relative distance of the agents satisfies the following scaling behavior (for all $\beta_h$ values in the interval $[0.01, 1]$, also note that it is not applicable for R2 regime)
\begin{equation}
    d_{\beta_d,\beta_h}^{\text{R1}}(t)=C(\beta_h)\beta_d^{-b}F\left(\beta_d^a t\right),
\label{Eq:data_collapse}
\end{equation}
where $a$ and $b$ are their corresponding exponents, $C(\beta_h)$ is a smooth function of $\beta_h$ and $F$ is a universal function with the asymptotic behavior $\lim_{x\rightarrow 0}F(x)\propto x^{b/a}$ and $\lim_{x\rightarrow \infty}F(x)= const$. These exponents are interestingly more or less independent of $\beta_h$ for $\beta_h\in [0.01, 1]$, being fixed at $a=2.00\pm 0.05$ and $b=1.00\pm 0.05$. This shows that $d(t)_{\text{small times}}\propto t^z$ where the dynamic exponent $z=0.51\pm0.02$ lies pretty within the normal diffusion regime. We notice that this behavior cannot be valid for (or simply extrapolated to) much larger $\beta_h$s where one expects the SAW regime with $z_{\text{SAW}}=\frac{3}{4}$.

\begin{figure*}
    \centering
    \includegraphics[width=0.9\textwidth]{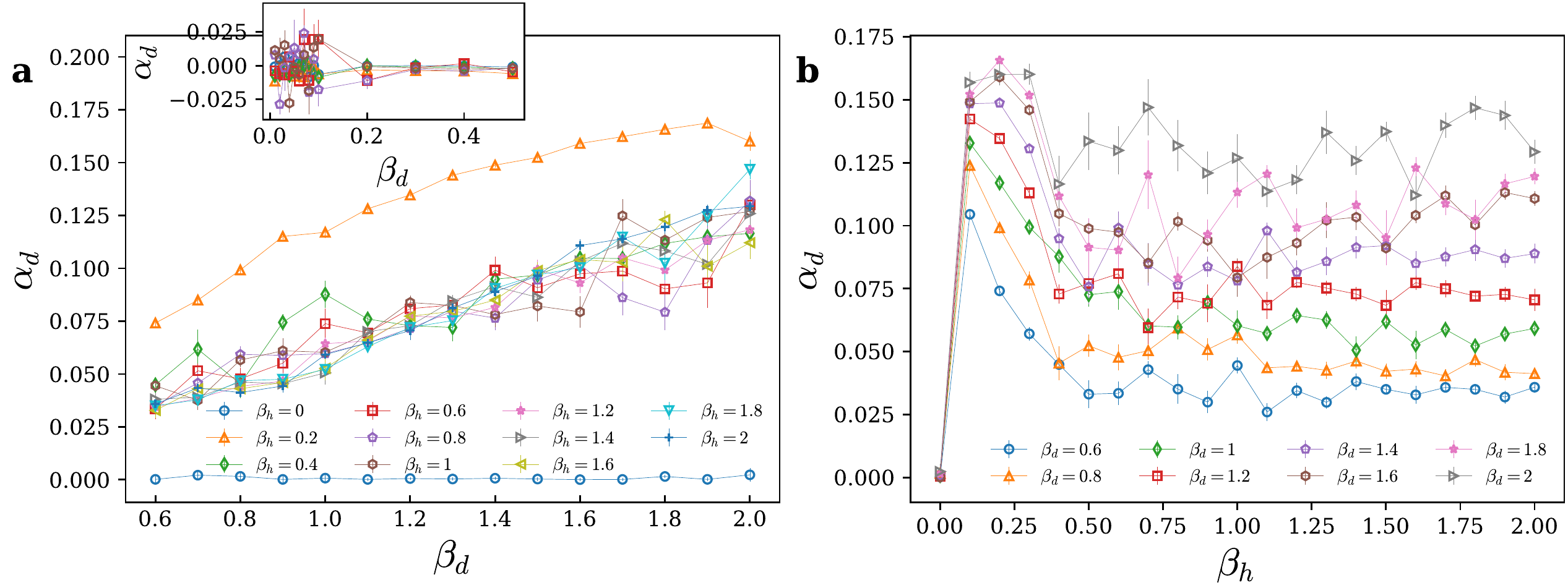}
    \caption{ The exponents $\alpha_d$ in terms of $\beta_d$ for $0.5 < \beta_d < 2$, and $0 < \beta_d \le 0.5$ are shown in panel-a-main, and panel-a-inset, respectively. In two mentioned regimes, R1 and R2, we see different behaviors for $ d(t) = |\textbf{r}_1-\textbf{r}_2| $. In the first regime, $d(t)$ reaches to a plateau region. But in the second regime, we see $d(t)$ follows the power law in terms of $\log t$. In panel-b, the exponent $\alpha_d$ is plotted in terms of $\beta_h$, which roughly remains fixed for almost large $\beta_h$.} 
\label{fig:alpha_d}
\end{figure*}

\begin{figure}
    \centering
    \includegraphics[width=0.45\textwidth]{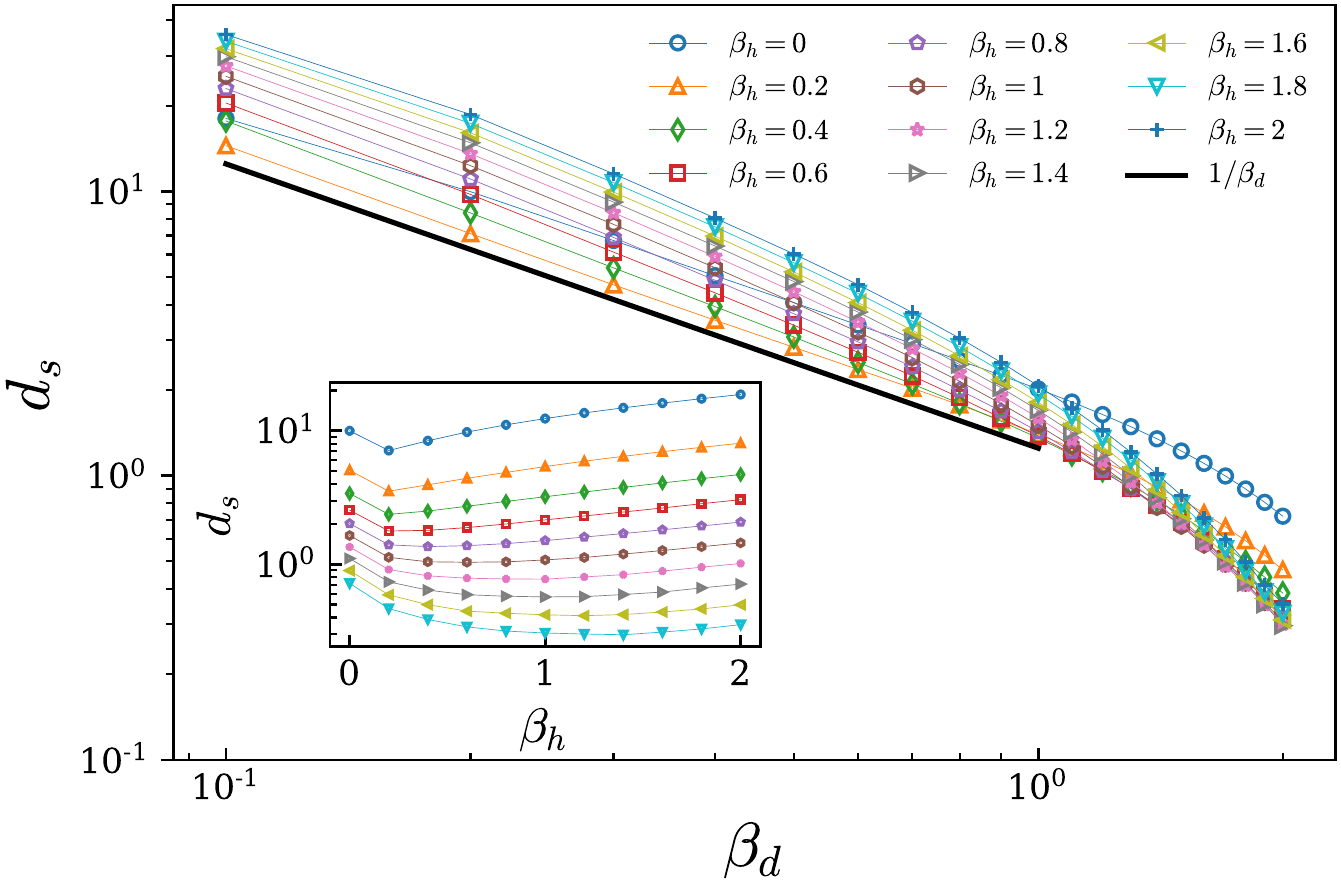}
    \caption{ The amount of $d$ at $t=10^3$ in terms of $\beta_d$ for various values of $\beta_h$ is shown in main panel. For small values of $\beta_d$, $d(t)$ reaches to $d_s$ and all curves follow the scaling relation $d_s \propto 1 / \beta_d$, which is consistent with Eq.~\eqref{Eq:mean}. The change of $d_s$ with respect to $\beta_h$ is negligibly small, which is shown in inset.}
\label{fig:d_saturation}
\end{figure}

To monitor the differences of the R1 and R2 regimes, we show the $\alpha_d$ exponent in terms of $\beta_d$ and $\beta_h$ in Figs.~\ref{fig:alpha_d}a and~\ref{fig:alpha_d}b, respectively. $\alpha_d$ is almost zero for $\beta_d\lesssim 0.6$ (inset of Fig.~\ref{fig:alpha_d}a) as expected from the definition of the R1 regime and grows more or less linearly by increasing $\beta_d$ starting from $\beta_d\approx 0.6$ consistent with the above claim ($\beta_h=0$ is an exception for which $\alpha_d$ is almost zero everywhere). The dependence of $\alpha_d$ to $\beta_h$ is low (Fig.~\ref{fig:alpha_d}b), as can also be seen in the other observables like Fig.~\ref{fig:d_saturation} where $d_s$ is the amount of $d$ at a fixed time $t=10^3$ after the saturation is established. The latter figure shows that the change of $d_s$ with respect to $\beta_h$ is negligibly small. For small $\beta_d$ values, the graphs are fitted with the relation $d_s\propto 1 / \beta_d $, which is true for all $\beta_h$ values considered in this work (in agreement with Eqs.~\ref{Eq:mean} and~\ref{Eq:data_collapse}), while some deviations are observed for large $\beta_d$ values.\\

The problem of random walkers in two dimensions can readily be mapped to a (2+1)-dimensional growth by considering the statistics of the height produced by the amount of debris left by random walkers at each site. From this point of view, the roughness is arguably the significant quantity that identifies the system's universality class. It is defined by
\begin{equation}
w^2=\left\langle \overline{\left(h(\textbf{r})-\bar{h} \right)^2} \right\rangle,
\end{equation}
where the over line represents the spatial average $\overline{O}\equiv \frac{1}{m(t)}\sum_{x,y\in \Gamma(t)} O(x,y)$, the $\left\langle \cdots \right\rangle $ is the ensemble average. The variables $m(t)=\sum_i \Theta(h_i(t)-1)$ and $\Gamma(t)$ are the number of occupied sites and the set of all occupied sites respectively. Also, $\Theta$ is the step function defined by $\Theta(x)=1$ for $x\ge 0$ and zero otherwise. We found a same crossover point in terms of $t$, so that for $\beta_d\ll \beta_d^*$ (R1 regime) both $ h $ ($ = \langle \bar{h} \rangle$) and $w$ grow with time in a power-law fashion, while for $\beta_d\gg \beta_d^*$ (R2 regime) both of them grow with a power of $\log t$. For $\beta_d\approxeq \beta_d^*$ both behaviors are observed, one for small time scales and another for large times. 

\begin{figure*}
    \centering
    \includegraphics[width=0.9\textwidth]{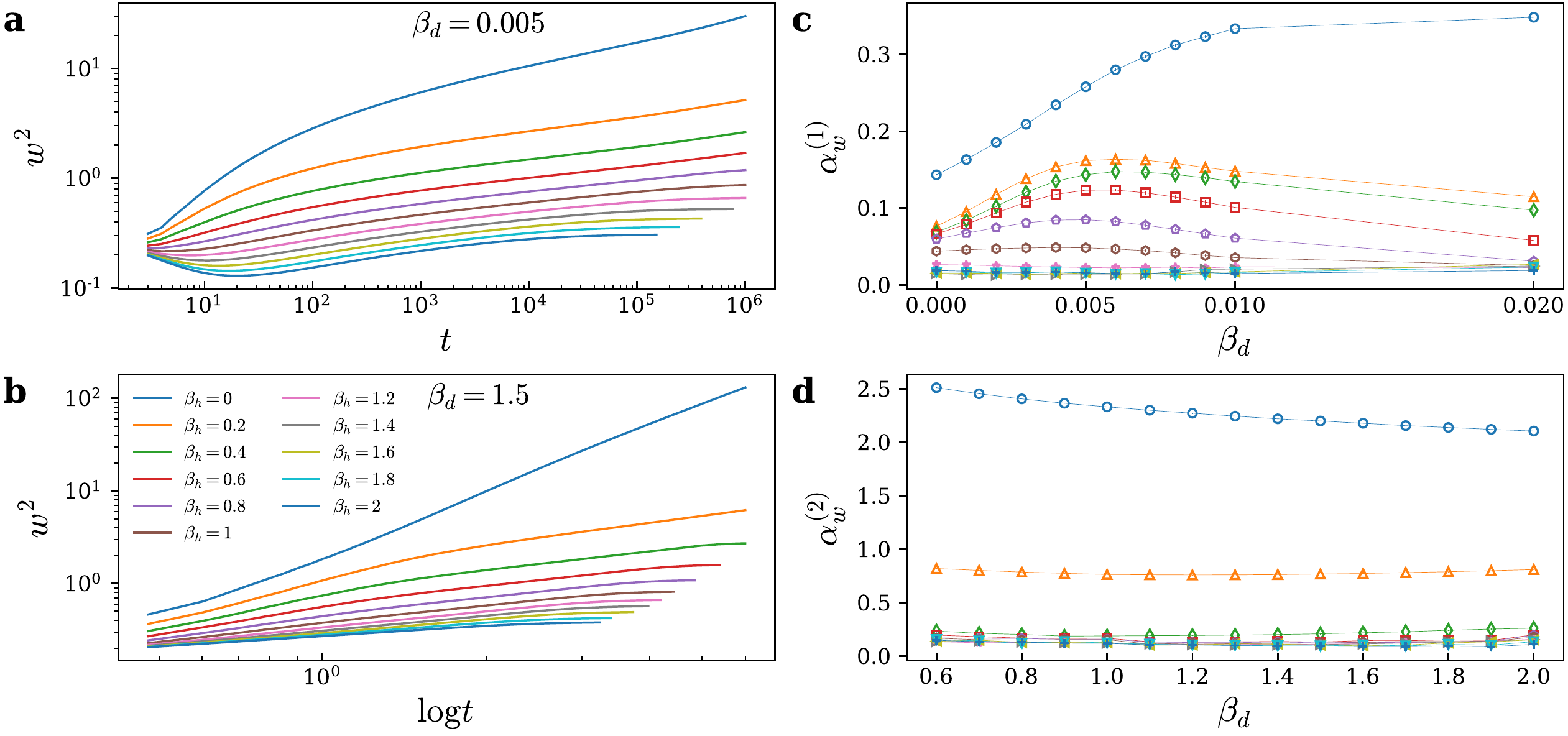}
    \caption{ Log-log plot of $ w^2$ in terms of $t$ for $\beta_d=0.005$ is plotted in panel-a. This $\beta_d$ lives in R1 regime, in which the width is proportional to $t^{\alpha_w^{(1)}}$. In panel-b, the log-log plot of width is shown in terms of $\log t$ for $\beta_d=1.5$. This $\beta_d$ belongs to R2 regime, in which $w^2(t) \propto (\log t)^{\alpha_w^{(2)}} $. In panel-c, and -d, the exponents $\alpha_w^{(1)}$, and $\alpha_w^{(2)}$ are depicted in terms of $\beta_d$ in the regions $\beta_d \ll 0.2$ and $\beta_d \gg 0.2$ for various amount of $\beta_h$, respectively. }
\label{fig:w_and_alpha}
\end{figure*}

Two extreme cases (R1 and R2 regimes) for $w$ have been shown in Fig.~\ref{fig:w_and_alpha} (the results for $ h $ is quite similar to $w$ which can be realized from scaling arguments). Fig.~\ref{fig:w_and_alpha}a and Fig.~\ref{fig:w_and_alpha}b are $w^2$ for $\beta_d=0.005$ (in R1 regime) and $\beta_d=1.5$ (R2 regime) respectively, from which we see that the log-log plot of roughness is linear with time in the R1 regime, whereas it is linear with respect to $\log t$ in the R2 regime. The quantities shown in~\ref{fig:w_and_alpha}c and ~\ref{fig:w_and_alpha}d are the corresponding exponents in long times defined by
\begin{equation}
w^2_{\text{R1}}\propto t^{\alpha^{(1)}_w}, \quad w^2_{\text{R2}}\propto \left(\log t \right)^{\alpha^{(2)}_w}.
\end{equation}
We see that, $\alpha_w^{(1)}$ after a small increase shows a decreasing behavior in terms of $\beta_d$ (it decreases with $\beta_h$), while $\alpha_h^{(2)}$ is more or less constant. Both exponents decrease with $\beta_h$ for all $\beta_d$ values. $\alpha_w^{(1)}$ is not a monotonic function in terms of $\beta_d$, showing a maximum at $\beta_d\approx 0.005$, while it decreases with $\beta_h$ confirming that it is entering a logarithmic regime. $\alpha_w^{(2)}$ is almost robust against $\beta_d$ and decrease with $\beta_h$. Note that our definition of spatial averaging differs from the one used in \cite{grassberger2017self} (here the average is over the occupied sites), so that our exponents cannot be compared with that paper. We are not sure whether this behavior of the roughness remains unchanged as $t\rightarrow\infty$ or it enters a stationary regime, for which a very larger scale simulations are needed. The second way out of this problem might be to consider the random walkers on a finite lattice, which is beyond this paper.

\section{fractal properties}\label{SEC:FD}

\begin{figure*}
    \centering
    \includegraphics[width=0.9\textwidth]{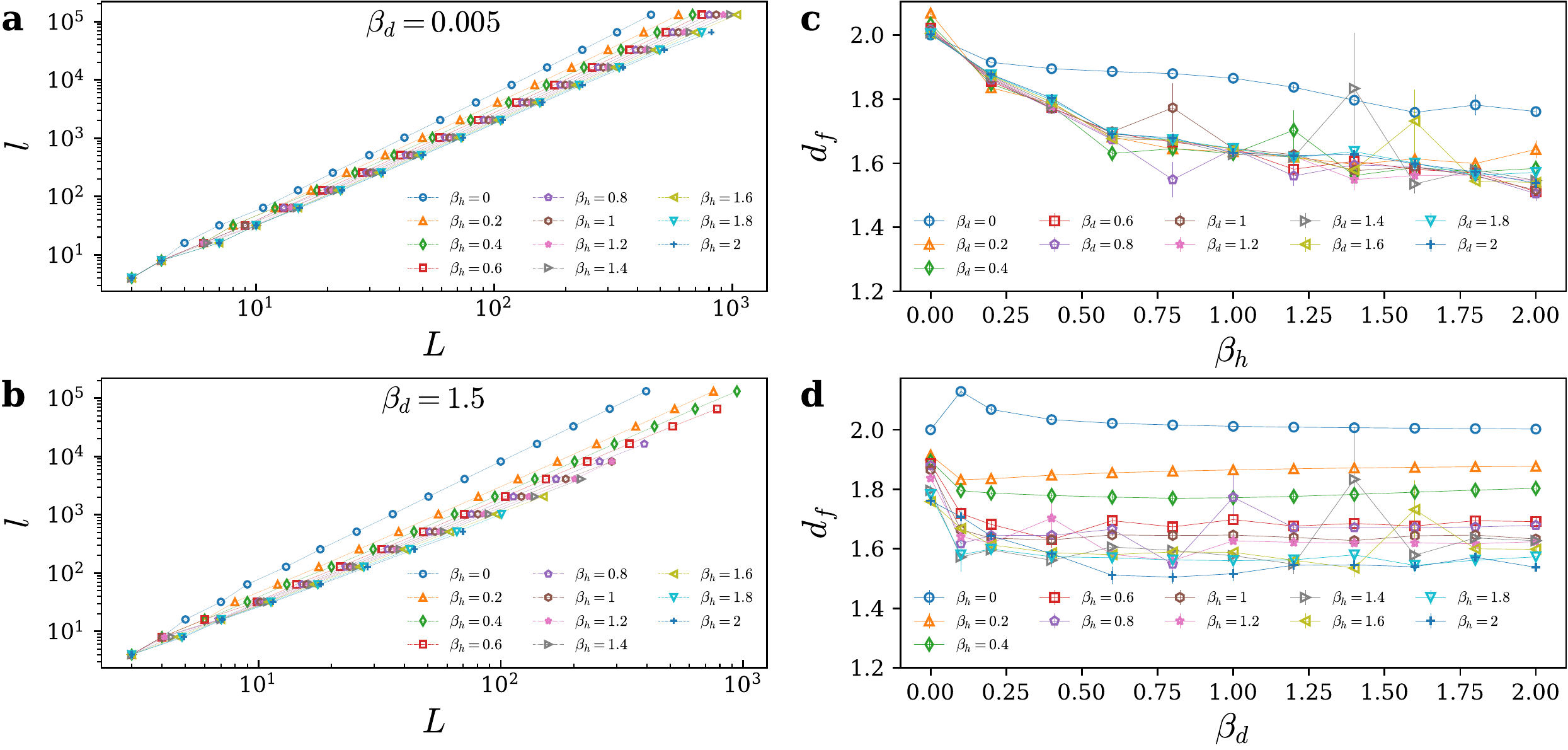}
    \caption{ Log-log plot of trace length of the random walker ($l$) in terms of $ L $ (lateral size of minimal square containing whole trace) for $\beta_d=0.005$ and $\beta_d=1.5$ are shown in panel-a and -b, respectively. The slope of each curve yields the fractal dimension $d_f$ of random walker traces.
    The corresponding fractal dimension calculated using the sandbox method in terms of $\beta_h$ and $\beta_d$ are shown in panel-c and -d, respectively. In panel-d, one can see that for $\beta_d \lesssim 0.2$, the value of $d_f$ changes abruptly. But for $\beta_d \gtrsim 0.2$, the value of $\alpha_{\theta}$ almost remains fixed.} 
\label{fig:sandbox_df}
\end{figure*}

The comparison of the fractal properties of the random walker traces with the other known exact results reveals the properties of the model. It especially helps much to understand the nature of the crossover. We have tested various definitions for the fractal dimension (FD), namely sandbox FD, box counting FD and the scaling relation between mass-gyration radius. The best one which fits best to the traces in our model is sandbox FD, for a description of the model see \cite{najafi2021self}. Briefly, one considers the traces for one random walker up to time $t$ with length $l(t)$ and enclose it with a minimal square with an edge length $L$. The scaling relation between $l$ and $L$ gives us the FD, $l(t) \sim L(t) ^{d_f} $.
In Fig.~\ref{fig:sandbox_df}a and \ref{fig:sandbox_df}b, this relation is shown for the R1 and R2 regimes respectively. The exponent $d_f$ can be considered as the dynamical mass fractal dimension since the quantities are time dependent. We calculate the effective $d_f$ in terms of $\beta_h$ and $\beta_d$ (Fig.~\ref{fig:sandbox_df}c and d respectively).
The fractal dimension of $\beta_d=\beta_h=0$ is almost $2$ as expected for 2D uncorrelated random walks. Note also that the other extreme is $\beta_h\rightarrow\infty$ for which $d_f^{\text{SAW}}=\frac{4}{3}$. From Fig.~\ref{fig:sandbox_df}c, we see that FD decreases as $\beta_h$ increases (showing that the traces become sparse), approaching this value. Fig.~\ref{fig:sandbox_df}d tells us that for $\beta_d>\beta_d^*$, FD becomes almost constant in terms of $\beta_d$.

\begin{figure*}
    \includegraphics[width=160mm]{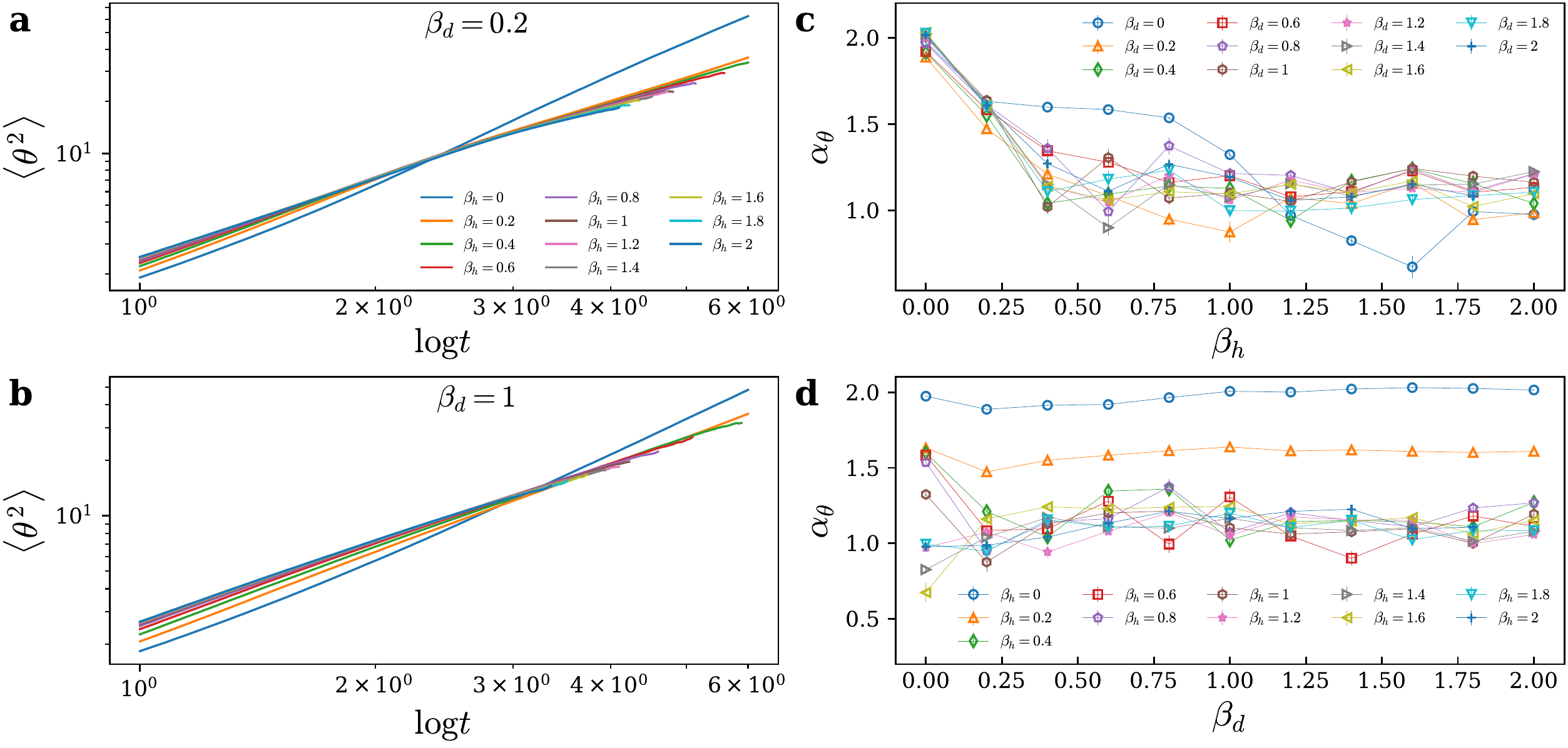}
    \caption{Log-log plot of the variance of the winding angle $\langle \theta^2 \rangle $ in terms of $\log t$ for $\beta_d=0.2$ and $\beta_d=1$ are shown in panel-a, and -b, respectively. The panel-a and -b have the same legend. The slope of each curve in the significant times gives $\alpha_\theta$. It seems that all curves in each panel meet together at a specific time say $t^*$. We will show the dependency of $t_*$ to $\beta_d$ in Fig.~\ref{fig:tstar_theta}. The corresponding exponent $\alpha_{\theta}$ in the long time limit in terms of $\beta_h$ and $\beta_d$ are plotted in panel-c and -d, respectively. For $\beta_d \lesssim 0.2$ in panel-d, one can see that the value of $d_f$ changes abruptly. But for $\beta_d \gtrsim 0.2$, the value of $\alpha_{\theta}$ almost remains fixed.}
\label{fig:theta_and_alpha}
\end{figure*}

Now we are in the position to test the statistics of the winding angle defined in Fig.~\ref{fig:winding_angle_def}. Its variance $\langle \theta^2 \rangle $ is given in Eq.~\eqref{Eq:winding} for $\beta_d=\beta_h=0$. This function is linear in the log-log scale plot in terms of $\log t$ shown in Fig.~\ref{fig:theta_and_alpha}a and \ref{fig:theta_and_alpha}b with the exponents in large time scales given in Fig.~\ref{fig:theta_and_alpha}c and d. These results confirm that the Eq.~\eqref{Eq:winding} is applicable for all cases with generalized exponents, i.e., it should be generalized to
\begin{equation}
  \left\langle \theta^{2}\right\rangle =A \left( \log t\right)^{\alpha_{\theta}(\beta_d,\beta_h)} \quad \alpha_{\theta}(0,0)=2,
\label{Eq:winding2}
\end{equation}
where $A$ is a non-universal constant. From Fig.~\ref{fig:theta_and_alpha}c and \ref{fig:theta_and_alpha}d, one observes that $\alpha_{\theta}$ changes from 2, to the lower values. It is hard to decide whether this exponent is identical for all $\beta_d$ and $\beta_h$ values when $\beta_h$ is high enough. Roughly speaking, $\alpha_{\theta}$ changes from $2$ for small $\beta_h$ and $\beta_d$ values to $0.9<\alpha_{\theta}(\beta_d,\beta_h)<1.2$ for large $\beta_h$ and $\beta_d$.

\begin{figure}
  \includegraphics[width=75mm]{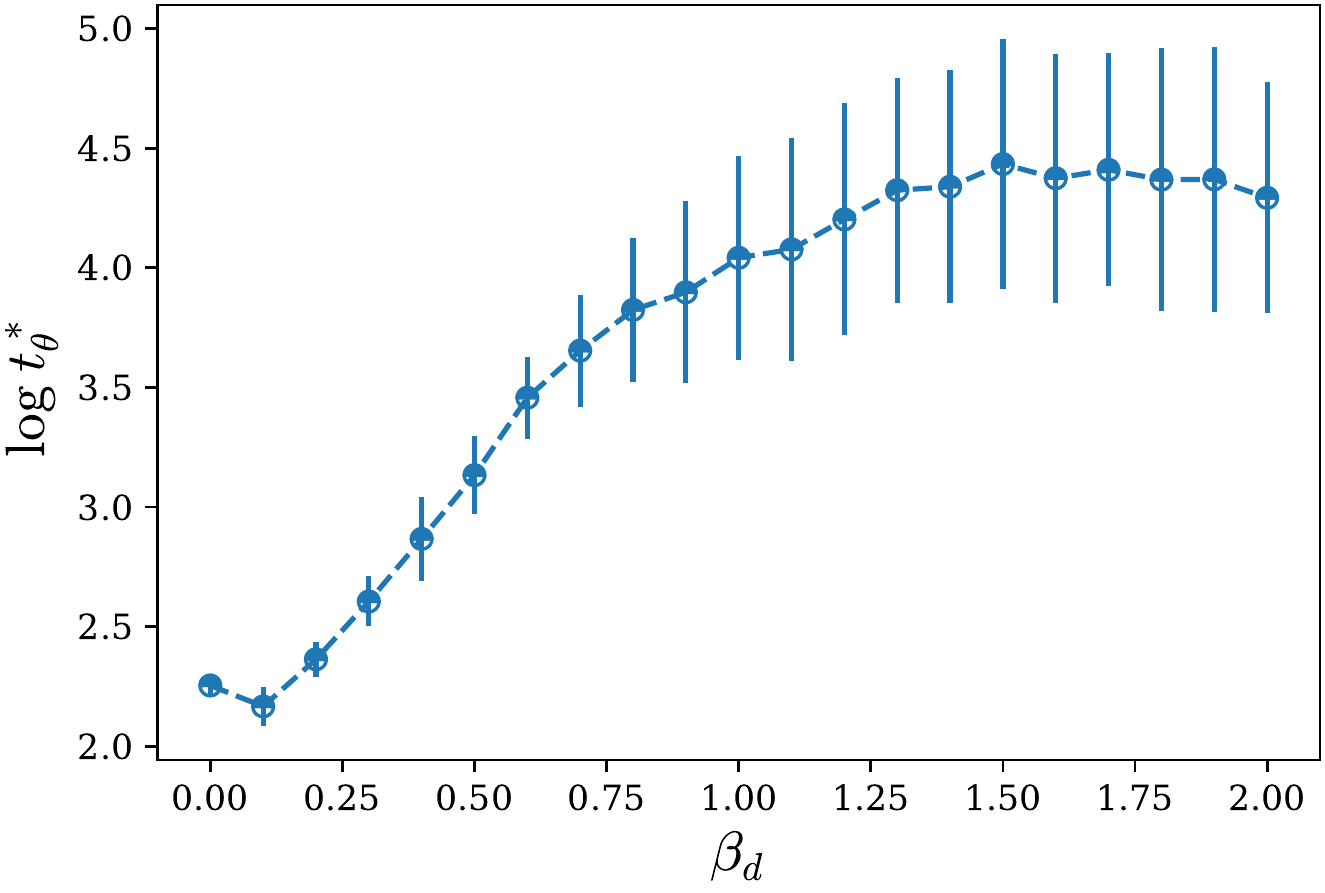}
  \caption{The crossover time $t^*_{\theta}$ related to $\langle \theta^{2}\rangle$ curves (see Eq.~\eqref{Eq:winding3} and Fig.~\ref{fig:theta_and_alpha}) in terms of $\beta_d$ is shown. It seems that for $\beta_d < 0.2$, the value of $t^*$ decreases. But for $\beta_d > 0.2$, it increases and finally for large $\beta_d$s greater than 0.5, the values of $t^*_{\theta}$ roughly remain fixed within the error bar. }
\label{fig:tstar_theta}
\end{figure}

It is interestingly seen that there is a universal point $t^*$ where for a fixed $\beta_d$, all the graphs for various $\beta_h$s meet each other. The slope of the graphs is different on two sides of this point, so that $t^*_{\theta}$ is served as the crossover point between small and large scale behaviors. We calculate $\alpha_{\theta}$ by extracting the slopes on the right-hand side of this point, i.e., large time scales. Using this fact, one can conclude that $A$ has no choice by $A= C / ( \log t^*_{\theta})^{\alpha_{\theta}(\beta_d,\beta_h)}$, where $C$ is a non-universal constant, reading
\begin{equation}
    \left\langle \theta^{2}\right\rangle = C \left( \frac{\log t}{\log t^*_{\theta}}\right) ^{\alpha_{\theta}(\beta_d,\beta_h)}.
\label{Eq:winding3}
\end{equation}
To calculate $t^*_{\theta}$, we identify the crossing point of each two curves corresponding to different values of $\beta_h$. The reported $t^*_{\theta}$ is the average value and the error bar is the variance of it. Figure~\ref{fig:theta_and_alpha}e shows $\log t^*_{\theta}$ in terms of $\beta_d$, in which we see that it changes behavior when one crosses from the R1 regime to R2 regime, i.e. in R1 it is an increasing function of $\beta_d$ while for the R2 regime it almost saturates to a constant.

\section{Conclusion}
\noindent
This paper was devoted to the analysis of a biped spider walk, i.e., two correlated self-repelling random walkers, which is realized by dropping debris in the lattice points that are visited. The motion of random walkers is controlled by two external parameters $(\beta_d,\beta_h)$ where $\beta_d$ captures the tendency of the random walkers to each other, and $\beta_h$ controls the possibility that a random walker steps on a site with debris height $h$. Our study uncovers the fact that there is a crossover point in which random walkers change from uncorrelated random walks to a new regime that is characterized in this paper in detail. In the new regime, the system is in the super diffusion phase with some diffusion exponents higher than $\frac{1}{2}$. By analyzing the diffusion exponent and also the fractal dimension of the random walker traces, we showed that the new regime although exhibiting properties partially similar to self-avoiding walks, represents new features. The first regime which is identified by a crossover region $\beta_d\lesssim\beta_d^*\in [0.2-0.5]$ ($\beta_d^*$ is a crossover point) is called R1 regime, whereas the other regime is R2.

In the long time limit, the random walkers stay in a finite equilibrium distance in the R1 regime, while they tend to each other in R2 regime. This tendency is described by a power-law decay of $d=|\textbf{r}_1-\textbf{r}_2|$ in terms of logarithm of time. The decay is faster for lower $\beta_d$s (Fig.~\ref{fig:alpha_d}a), while it is not very sensitive to $\beta_h$ (for not very small $\beta_h$s).\\
A similar crossover is seen for the sandbox fractal dimension of random walker traces, which is expected to become uncorrelated random walk $d_f^{\text{URW}} = 2$ as $(\beta_d,\beta_h)\rightarrow (0,0)$, and SAW as $(\beta_d,\beta_h)\rightarrow (0,\infty)$ with $d_f^{\text{SAW}}=\frac{4}{3}$. We observed that the traces become more sparse (more self-avoiding) as $\beta_h$ increases, i.e. $d_f$, starting from $2$ (the uncorrelated case) decreases with $\beta_h$ saturating to a value. For $\beta_d\ne 0$ this final fractal dimension is almost independent of $\beta_d$, e.g., it is $1.6\pm 0.1$ for $\beta_h=2.0$.

In the analysis of the winding angle, two important facts were found: 1- the variance of the winding angle regarding Eq.~\eqref{Eq:winding} with different exponent (Eq.~\eqref{Eq:winding3}) with an exponent which is not effectively $\beta_d$-dependent, and is pretty sensitive to $\beta_h$ (transiting from $2.0\pm 0.1$ to $1.1\pm 0.2$), 2- There is a crossover time $t^*_{\theta}$ where for fixed $\beta_d$ all the $\beta_h$ graphs meet each other almost in a same point. We observed that the slopes of the graphs before and after this point are slightly different (Fig.~\ref{fig:theta_and_alpha}).

\noindent
{\bf Acknowledgments.}
H.D.N. supported by a KIAS Individual Grant PG066902 at Korea Institute for Advanced Study. H.P. supported by a KIAS Individual Grant PG013604 at Korea Institute for Advanced Study. This work is supported by the Center for Advanced Computation at KIAS.

\bibliography{refs} 

\end{document}